\documentstyle[preprint,aps,epsfig]{revtex}

\begin{document}

\title{Probing neutrino mass hierarchies and $\phi_{13}$ with supernova neutrinos}

\author{Shao-Hsuan Chiu\thanks{schiu@mail.cgu.edu.tw}}
\address{Physics Group, C.G.E., Chang Gung University, 
Kwei-Shan 333, Taiwan}

\author{T. K. Kuo\thanks{tkkuo@physics.purdue.edu}}
\address{Department of Physics, Purdue University, West Lafayette, IN 47907}

\maketitle

\newif\iftightenlines\tightenlinesfalse
\tightenlines\tightenlinestrue

\begin{abstract}

We investigate the feasibility of probing the neutrino mass hierarchy
and the mixing angle $\phi_{13}$ 
with the neutrino burst from a future supernova.
An inverse power-law density $\rho \sim r^{n} $ with varying $n$
is adopted in the analysis as the density profile of a typical 
core-collapse supernova.  
The survival probabilities of 
$\nu_{e}$ and $\bar{\nu}_{e}$ are shown to reduce to 
two-dimensional functions of $n$ and $\phi_{13}$.  
It is found that in the $n-\sin^{2} \phi_{13}$ parameter
space, the 3D plots of the probability functions exhibit highly
non-trivial structures that are sensitive to the mass hierarchy, 
the mixing angle $\phi_{13}$,
and the value of $n$.  The conditions that lead to 
observable differences in the 3D plots are established. 
With the uncertainty of $n$ considered, 
a qualitative analysis of the Earth matter effect is also included.

\end{abstract}


\pagenumbering{arabic}



\section{Introduction}

A better knowledge of the neutrino property is regarded
as one of the crucial keys in searching for new physics beyond the
standard model.
Recent experiments of neutrino oscillation have been able to uncover
part of the neutrino properties, such as the neutrino mixing angles
and the mass squared differences responsible for the
solar neutrino problem and the atmospheric 
neutrino deficit~\cite{sk:04,sno:01,sk:2004,kam:03,k2k:03}.
However, there exists an ambiguity in the sign of the squared difference 
$\Delta m^{2}_{32} \equiv m^{2}_{3}-m^{2}_{2}$ involved in 
the oscillation of atmospheric neutrinos.   
The two scenarios are referred to as the normal 
($m_{3} \gg m_{2}, m_{1}$) and the
inverted ($m_{2}, m_{1} \gg m_{3}$) mass hierarchies.
Furthermore, although an upper bound on the mixing angle
$\phi_{13}$ is established by 
$\sin^{2}\phi_{13} <0.02$~\cite{chooz:99,palo:00}, a definite determination of
this angle has not been achieved yet.

The rich physical content of a core-collapse supernova
makes the supernova neutrino one of the most promising tools for 
the study of  
unknown neutrino properties and the supernova mechanism~\cite{raffelt}.
The supernova neutrinos are unique in that both neutrinos and antineutrinos
are produced at very high densities and high temperatures before
propagating through matter of varying densities.
Due to the wide range of matter density in a supernova,
the neutrinos may go through two or even three
(if the regeneration effect due to the Earth matter occurs)
separate flavor conversions before reaching the terrestrial detectors.
Furthermore, the matter-enhanced oscillations~\cite{w,ms} in a core-collapse 
supernova lead to a striking feature that a small variation
of the mixing angle $\phi_{13}$ can significantly alter the neutrino spectra. 
For supernova neutrinos, the main physical consequence 
arising from the ambiguity of the mass hierarchy is that 
both the higher and the lower level crossings occur
in the $\nu$ sector if the mass hierarchy is normal, while the higher
crossing occurs in the $\bar{\nu}$ sector and the lower crossing
occurs in the $\nu$ sector if the mass hierarchy is inverted.

The future galactic supernova is capable of 
inducing roughly $10^{4}$ neutrino events at the terrestrial detectors,
and is expected to provide a much better statistics 
than the SN1987A~\cite{SN1987A:87} did.
This promising characteristic has motivated a wealth of 
discussions on how the neutrino fluxes
from a supernova can facilitate the search of the unknown neutrino
properties~\cite{dighe:00,fogli:02,luna:03,dighe:03,bandy:03,dav:05,chiu:00}.
As generally realized, the main difficulty in 
extracting information from the supernova neutrinos arises from the poorly 
known exploding mechanism.   Incomplete knowledge of the supernova leads to, 
among others, an uncertainty in the density profile of a supernova.

The supernova neutrinos are usually assumed to propagate
outward through an inverse power-law matter density, $\rho \sim r^{n}$,
with $n=-3$.
However, due to lack of statistically significant real data, 
there is no clear evidence showing that
the density distribution $\rho \sim r^{-3}$ provides a reliable connection
between the dynamics of flavor conversion and the expected
neutrino events at the detectors.  In addition, the shock propagation 
in a supernova~\cite{fogli:03} represents a time-dependent 
disturbance to the matter density and causes 
a sizable effect to the neutrino flavor conversion.
Since only the matter density near a resonance point is 
relevant to the flavor conversion and
any local deviation from $n=-3$ cannot be ruled out, 
the profile $\rho \sim r^{-3}$ should not 
be considered as a satisfactory description to the density shape
for the purpose of extracting neutrino properties from the 
observation of supernova neutrinos.

In this work, possible consequences resulting from
variation of the density profile are examined.
The aim is to analyze the neutrino survival probabilities
and to study how the uncertainty in $n$ would affect the determination of
the mixing angle $\phi_{13}$ and the mass hierarchy.
This paper is organized as follows.  In Section II, we outline
the $n$-dependent formulation of the survival probabilities 
for $\nu_{e}$ and $\bar{\nu}_{e}$.  
In section III, parameters obtained from the solar, the atmospheric, and the
terrestrial experiments are taken as the input for constructing  
the 3D plots of the probability functions.  
The probability functions under both the normal and the inverted 
mass hierarchies are analyzed.  
In section IV, we discuss the feasibility of probing the neutrino
mass hierarchy and $\phi_{13}$ with the properties of the probability functions. 
In section V, the discussion is expanded to include 
the Earth matter effect.  We then summarize this work in section VI.

 
\section{Conversion probabilities}

The density profile of matter encountered 
by the propagating neutrinos plays a crucial role in the dynamics
of flavor conversion.  In the literature, 
the neutrino flavor conversion in media of various 
density profiles has been widely discussed. However, the exact solution
is obtained only for a few specific density distributions
: the linear, exponential, hyperbolic tangent, 
and the $1/r$ profiles.
It was suggested~\cite{kuo:39} that for an arbitrary inverse power-law density
$\rho \sim r^{n}$, an extra correction factor $F$ (a function
of $n$ and the mixing angle) 
can be supplemented to the standard Landau-Zener~\cite{Landau}
formulation of the level crossing to account for the 
effect due to deviation from a linear density profile.

With the extremely high electron number density in a supernova, 
the effective mixing angles in matter for the neutrino and the 
antineutrino become
$\phi^{m}_{13} \approx \pi/2$, $\theta^{m}_{12} \approx \pi/2$ and 
$\bar{\phi}^{m}_{13} \approx 0$, $\bar{\theta}^{m}_{12} \approx 0$,
respectively.  
Using the standard parametrization of the neutrino mixing matrix: 
$U^{2}_{e1}=\cos^{2}\phi_{13} \cos^{2}\theta_{12}$,
$U^{2}_{e2}=\cos^{2}\phi_{13} \sin^{2}\theta_{12}$,
and $U^{2}_{e3}=\sin^{2}\phi_{13}$,
the survival probabilities for $\nu_{e}$ and $\bar{\nu}_{e}$ 
can be written respectively as~\cite{kuo:61,fogli:02}

\begin{equation}\label{eq:pe}
       P_{nor}=U_{e1}^{2} P_{l} P_{h} + U_{e2}^{2} (1-P_{l}) P_{h} +
                 U_{e3}^{2} (1-P_{h}),
\end{equation} 
   
\begin{equation}\label{eq:pa}
   \bar{P}_{nor}=U_{e1}^{2} (1-\bar{P}_{l}) + U_{e2}^{2} \bar{P}_{l},
\end{equation} 
for the normal hierarchy and

\begin{equation}\label{eq:pei}
    P_{inv}=U_{e2}^{2} (1-P_{l}) + U_{e1}^{2} P_{l},
\end{equation}
   
\begin{equation}\label{eq:pai}
    \bar{P}_{inv}=U_{e2}^{2} \bar{P}_{l} \bar{P}_{h}+
                  U_{e1}^{2} (1-\bar{P}_{l}) \bar{P}_{h} + 
                 U_{e3}^{2} (1-\bar{P}_{h}),
\end{equation}
for the inverted hierarchy.  Note that $P_{h}$ ($\bar{P}_{h}$) 
and $P_{l}$ ($\bar{P}_{l}$)
represent the higher and the lower level crossing probabilities for 
$\nu_{e}$ ($\bar{\nu}_{e}$), respectively.  
For arbitrary density profile and mixing angle,
the Landau-Zener formula is modified as 

\begin{equation}\label{eq:pc} 
 P_{h,l} = \frac{\exp(-\frac{\pi}{2} \gamma_{h,l} F_{h,l})-
 \exp(-\frac{\pi}{2} \gamma_{h,l} \frac{F_{h,l}}{\sin^{2} \Theta_{ij}})}
 {1-\exp(-\frac{\pi}{2} \gamma_{h,l} \frac{F_{h,l}}{\sin^{2} \Theta_{ij}})},
   \end{equation}
where 
$\gamma_{h,l}$ are the adiabaticity parameters,
$\Theta_{ij}$ are the mixing angles between the $i$th and the $j$th
mass eigenstates,
and $F_{h,l}$ are the correction factors to a non-linear profile.
Note that $P_{h} = \bar{P}_{h}$, and that $\bar{P}_{l}$ 
can be obtained directly from $P_{l}$ by replacing $\Theta_{ij}$ with 
$\pi/2-\Theta_{ij}$.

For a typical core-collapse supernova, the electron number density can be written as 
$N_{e}=(\frac{Y_{e}}{m_{n}}) c r^{n}$, 
where $Y_{e}$ is the electron number per baryon, $m_{n}$ is the baryon mass,
and $c$ is a constant  
representing the scale of the 
density\footnote{The value of $c$ varies very weakly with $r$ over the range 
$10^{12} g/cm^{3} < \rho < 10^{-5} g/cm^{3}$~\cite{kuo:298}.}.
The adiabaticity parameter for this density profile has the general form
 
\begin{equation}\label{eq:gamma}
   \gamma_{h,l}=\frac{1}{2|n|}(\frac{\Delta m^2_{ij}}{E})^{1+\frac{1}{n}}
   (\frac{\sin^2 2\Theta_{ij}}{\cos 2\Theta_{ij}})
   (\frac{\cos 2\Theta_{ij}}
   {2\sqrt{2} G_{F}\frac{Y_{e}}{m_{n}} c})^{\frac{1}{n}},
\end{equation}
where $G_{F}$ is the Fermi constant, $E$ is the neutrino energy, and
$\Delta m^2_{ij} \equiv |m^{2}_{i}-m^{2}_{j}|$.
In the numerical calculation, it would be more convenient to 
write $F_{h}$ and $F_{l}$ as the Euler integral representation
of the hypergeometric function:
 
\begin{eqnarray}\label{eq:hp}
F_{h,l} & = &_{2}F_{1} (\frac{n-1}{2n},
\frac{2n-1}{2n};2;-\tan^2 2\Theta_{ij}) \nonumber \\
 & = & \frac{\Gamma (2)}{\Gamma (\frac{2n-1}{2n}) \Gamma(2-\frac{2n-1}{2n})}
\int_{0}^{1}  t^{(\frac{2n-1}{2n}-1)} (1-t)^{(2-\frac{2n-1}{2n}-1)}
[1-t(-\tan^2 2\Theta_{ij})]^{(\frac{n-1}{2n})} dt. 
\end{eqnarray}

The above expressions for $P_{h,l}$, $\gamma_{h,l}$,
and $F_{h,l}$ can be applied to an arbitrary profile and to both large or small
mixing angles.  
It was pointed out~\cite{friedland:01,valle:02} 
that there exists a subtlety in the physical meaning of resonance conversion: 
For the large mixing angle, the adiabaticity parameters,
$\gamma_{h}$ and $\gamma_{l}$,  
should each be calculated at the point of maximum violation of adiabaticity (PMVA)
instead of the point of resonance. 
Note that while the values of $\gamma_{h,l}$ depend on 
the locations where they are calculated, 
the values of $\gamma_{h} F_{h}$ and $\gamma_{l} F_{l}$ remain invariant. 
To simplify the calculation, 
we choose to evaluate $\gamma_{h} F_{h}$ and $\gamma_{l} F_{l}$
at the locations of resonance.


\section{Numerical analysis of $P$ and $\bar{P}$}


With the numerical input of $\Delta m^2_{21}$, $\Delta m^2_{32}$,
$\theta_{12}$, and $c$ (see, {\em e.g.}, 
ref.~\cite{luna:03} and the references therein),
the survival probability functions   
$P=P(\Delta m^2_{21}, \Delta m^2_{32}, \phi_{13}, \theta_{12}, E_{\nu}, n, c)$
and $\bar{P}=\bar{P}(\Delta m^2_{21}, 
\Delta m^2_{32}, \phi_{13}, \theta_{12}, E_{\bar{\nu}}, n, c)$
reduce to $P=P(E_{\nu}, n, \phi_{13})$ and 
$\bar{P}=\bar{P}(E_{\bar{\nu}}, n, \phi_{13})$, respectively.
The ambiguity of the mass hierarchy gives rise to four
distinct probability functions to be investigated: 
$P_{nor}$, $\bar{P}_{nor}$ for the normal hierarchy and 
$P_{inv}$, $\bar{P}_{inv}$ for the inverted hierarchy.
Their properties can be examined by a series of 3D plots
in the $E-\sin^{2}\phi_{13}$ parameter space.
As an illustration, we show only the 3D plots of
$P_{nor}$ for several $n$ values in Fig. 1.
From the 3D plots of $P_{nor}$, $\bar{P}_{nor}$, $P_{inv}$, 
and $\bar{P}_{inv}$, it can be concluded 
that all the probabilities exhibit no significant energy dependence
except for the low energy end.
This behavior implies that when the adiabaticity parameters in Eq. (6)
are calculated, the impact coming from the variation of
$n$ and $\phi_{13}$ dominate over that of the variation of energy in the typical
range, $E <10^{2}$ MeV.

Since the neutrino population is extremely small at the 
low energy end of the spectrum,
it would be convenient to simply take the average energies, 
{\em e.g.}, $\langle E_{\nu}\rangle=12$ MeV and 
$\langle E_{\bar{\nu}} \rangle=15$ MeV, in the numerical calculation.
This approximation further reduces the probabilities to functions
of only $n$ and $\phi_{13}$.  
The survival probabilities for $\nu_{e}$
and $\bar{\nu}_{e}$ can be plotted in the $n-\sin^{2}\phi_{13}$ space,
as shown in Fig. 2 and Fig. 3, respectively. 
It is seen that if $n < -6$,
the values of all the probability functions approach a constant $\sim 0.6$
regardless of the mass hierarchy and the value of $\phi_{13}$. 
Thus, the mass hierarchies are indistinguishable and the information about
$\phi_{13}$ is lost if $n < -6$.  
As $n$ increases from $-6$, the probability functions 
of $\nu_{e}$ in Fig. 2 are seen to drop through a
transition near $n \sim -5$, while that of $\bar{\nu}_{e}$ in Fig. 3
jump through a transition near $n \sim -4.5$.
Furthermore, $P_{nor}$ and $\bar{P}_{inv}$ exhibit an extra non-trivial 
structure for $n > -4$. 
In the following discussion, we divide $n$ 
into three regions: $n < -6$, $-6 < n < -4$, and $n > -4$.
The probability functions for $\nu_{e}$ and $\bar{\nu}_{e}$
shall be discussed separately.

\subsection{$P_{nor}$ and $P_{inv}$}
We first note that the condition
$P_{nor} = P_{inv}$ is satisfied if the higher crossing 
is extremely non-adiabatic: $P_{h} \rightarrow 1$, as implied by
Eqs.~(\ref{eq:pe}) and~(\ref{eq:pei}).
Thus, $P_{nor}$ and $P_{inv}$ are indistinguishable if the values
of $n$ and $\sin^{2}\phi_{13}$ result in $P_{h} \rightarrow 1$,
which occurs in part of the $n-\sin^{2} \phi_{13}$ parameter space,
as can be seen by comparing Figs. 2(a) and 2(b).
To account for the $n$-dependent transition of both $P_{nor}$ and $P_{inv}$
in the range $-6< n <-4$, we note that the adiabatic parameter at the lower crossing 
takes the form
\begin{equation}\label{eq:power}
  \gamma_{l} \simeq \frac{0.43 \times 10^{-5}}{|n|}
  (0.39 \times 10^{-30})^{\frac{1}{n}},
\end{equation}
which yields $\gamma_{l} \ll 1$ and $P_{l} \simeq \cos^2 \theta_{12}$ 
(non-adiabatic transition) for $n < -6$.
On the other hand, the same expression leads to 
$\gamma_{l} \gg 1$ and $P_{l} \approx 0$ (adiabatic transition) when
$n > -4$.
It is clear that $\gamma_{l}$ goes through a transition from 
$\gamma_{l} \ll 1$ to $\gamma_{l} \gg 1$ as $n$ varies from $n \sim -6$
to $n \sim -4$.
A simple calculation shows that $P_{inv} \simeq P_{nor} \approx 0.6$ 
at $n=-6$, 
while $P_{inv} \simeq P_{nor} \approx 0.3$ at $n=-4$ if $\sin^{2}\phi_{13} < 10^{-3}$.  
This result implies that the uncertainty of $n$ between $n=-4$ and $n=-6$ could lead to 
a variation of the survival probability by a factor of two
and complicate the interpretation of the $\nu_{e}$ events.

The 3D plots of $P_{inv}$ and $P_{nor}$ in the $n-\sin^{2}\phi_{13}$ 
space become distinguishable for $n >-4$ if $P_{h} \neq 1$.
Note that for $n > -4$, $P_{l} \approx 0$ and $P_{nor}$ becomes
\begin{equation}\label{eq:nud}  
  P_{nor} \simeq \sin^2 \phi_{13} + P_{h} (\sin^2 \theta_{12}
   \cos^2 \phi_{13}-
  \sin^2 \phi_{13}),
\end{equation}
where $P_{h}$ is given by Eq.~(\ref{eq:pc}), 
and the adiabaticity parameter $\gamma_{h}$
in $P_{h}$ is given by Eq.~(\ref{eq:gamma}):  
\begin{equation}\label{eq:power-h}
  \gamma_{h} \simeq \frac{10^{-4}}{|n|}
  (37.6 \times 10^{-30} \times \cos 2\phi)^{\frac{1}{n}} 
  \frac{\sin^2 2\phi_{13}}{\cos 2\phi_{13}}.
\end{equation}  

For $n>-4$, the subtle dependence of $P_{nor}$ on $n$ and $\sin^{2}\phi_{13}$ 
can be seen clearly from Eq.~(\ref{eq:nud}) and Fig. 2.
For small $|n|$ and large $\phi_{13}$ 
(near the right corner of Fig. 2(a)), the higher level crossing
is adiabatic ($\gamma_{h} \gg 1$ and $P_{h} \approx 0$).
Thus, the first term 
in Eq.~(\ref{eq:nud}) dominates and $P_{nor} \simeq \sin^2 \phi_{13} \ll 1$.  
On the other hand, 
the higher level crossing becomes non-adiabatic 
($\gamma_{h} \ll 1$ and $P_{h} \approx 1$) 
for relatively larger $|n|$ and smaller $\phi_{13}$. 
Thus, the second term in Eq.~(\ref{eq:nud}) begins to dominate, and 
$P_{nor} \simeq \sin^2 \theta_{12} \approx 0.3$. 
Note that since the two values of $P_{h}$, 
$P_{h} \approx 1$ and $P_{h} \approx 0$, give rise to
the above two distinct values of $P_{nor}$ representing the two sides
of the fast transition area in Fig. 2(a), the condition
$P_{h}=1/2$ should reasonably describe the fast transition
of $P_{nor}$ for $n > -4$.
Furthermore, due to the smallness of the upper bound on $\phi_{13}$, 
the arguments in the numerator of Eq.~(\ref{eq:pc}) satisfy the relation
\begin{equation}
  \frac{\pi}{2} \gamma_{h} F_{h} \ll 
  \frac{\pi}{2} \frac{\gamma_{h} F_{h}}{\sin^{2} \phi_{13}},
\end{equation}
which implies that  
\begin{equation}
  \exp(-\frac{\pi}{2} \gamma_{h} F_{h}) \approx \frac{1}{2}
\end{equation}
at the narrow transition region.
The condition Eq. (12) leads to $G_{h} \approx 1$, where
\begin{equation}
  G_{h} \equiv \frac{\pi}{4(\ln2)}
  \frac{1}{|n|}(\frac{\Delta m^{2}_{32}}{E_{\nu}})
  (\frac{\frac{\Delta m^{2}_{32}}{E_{\nu}} \cos 2\phi_{13}}{c})^{\frac{1}{n}}
  (\frac{\sin^{2}2\phi_{13}}{\cos2\phi_{13}}) F_{h}.
\end{equation}
Since $F_{h} \sim 1$
in the region of interest ($n > -4$ and $\sin^{2}\phi_{13} < 10^{-2}$), 
the above condition can be approximated as

\begin{equation}\label{eq:g}
  G_{h}(n,\phi_{13}) \simeq \frac{2.3 \times 10^{-4}}{|n|}
  (37.6 \times 10^{-30} \times \cos 2\phi_{13})^{\frac{1}{n}} 
  \frac{\sin^{2} 2\phi}{\cos 2\phi_{13}} \approx 1.
\end{equation}
It can be shown that $P_{h} \approx 1$ if $G_{h}(n,\phi_{13}) <1$
and $P_{h} \approx 0$ if $G_{h}(n,\phi_{13}) >1$.
Take $n=-3$ for example, the sudden probability transition is located near
$\sin^{2}\phi_{13} \sim 10^{-5}$. Thus, 
$P \sim 0$ if $\sin^{2}\phi_{13} > 10^{-5}$, which is unique
to the normal mass hierarchy.  However, 
if $\sin^{2}\phi_{13} < 10^{-5}$, then $P \sim 0.3$ for both
the normal and the inverted mass hierarchies.  
This feature is clearly seen in Fig. 2.

Eq.~(\ref{eq:g}) suggests that a slight variation of the power 
may cause an ambiguity in the interpretation of $\phi_{13}$ and the mass hierarchy 
that are derived from the observation of neutrino events.
Note that although the numerical
values in Eq.~(\ref{eq:g}) vary with
the input parameters, the physical content remains unaltered.
We summarize this subsection as follows.
\begin{enumerate}
\item Given the input values of 
$\Delta m^2_{21}$, $\sin^2 \theta_{12}$, and $c$, 
it can be shown that $P_{l}$ is adiabatic ($\gamma_{l} \gg 1$) for 
$n > -4$ and non-adiabatic ($\gamma_{l} \ll 1$) for $n < -6$.
\item For the normal mass hierarchy, the two distinct 
values of $P$ due to $P_{h} \approx 1$ and $P_{h} \approx 0$ for $n > -4$
are separated by the condition Eq.~(\ref{eq:g}).
\item In principle, a direct observation of the $\nu_{e}$ events
could be used to distinguish the mass hierarchy if the values of 
$n$ and $\phi_{13}$ 
result in $P_{h} \neq 1$, as suggested by Figs. 2(a) and 2(b).
\end{enumerate}

\subsection{$\bar{P}_{nor}$ and $\bar{P}_{inv}$}

The survival probabilities of $\bar{\nu}_{e}$
are described by Eq.~(\ref{eq:pa}) and Eq.~(\ref{eq:pai}) 
for the normal and the inverted
mass hierarchies, respectively. 
The 3D plots of $\bar{P}_{nor}$ and $\bar{P}_{inv}$ are shown in Fig. 3.
Note that $P_{h} = \bar{P}_{h}$, and that $\bar{\gamma}_{l}$, $\bar{F}_{l}$, 
and $\bar{P}_{l}$ can be obtained respectively from $\gamma_{l}$,
$F_{l}$, and $P_{l}$ by the swap $\sin \theta_{12} \leftrightarrow
\cos \theta_{12}$.  Since $\bar{P}_{l} \simeq \sin^{2}\theta_{12} \approx 0.3$ 
and $\bar{P}_{h} \approx 1$ for $n < -6$, it follows that

\begin{equation}
 \bar{P}_{nor} = \bar{P}_{inv}
\simeq \cos^2 \phi_{13} \cos^2 \theta_{12} (1-\sin^2 \theta_{12})
+\cos^{2}\phi_{13} \sin^{4}\theta_{12} \approx 0.6. 
\end{equation}
This explains why the mass hierarchies are also indistinguishable from observing the
$\bar{\nu}_{e}$ events if $n < -6$.  In addition, the transition behavior of
$\bar{P}_{nor}$ and $\bar{P}_{inv}$ for $-6 < n < -4$ can be explained  
in the way similar to that of $P_{nor}$ and $P_{inv}$.

For $n > -4$, the lower level crossing becomes adiabatic: 
$\bar{P}_{l} \approx 0$. 
If the higher crossing remains non-adiabatic
($\bar{P}_{h} \approx 1$), it leads to $\bar{P}_{nor} \simeq \bar{P}_{inv}
\simeq \cos^2 \phi_{13} \cos^2 \theta_{12} \approx 0.7$,
which is slightly higher than that for $n < -6$.
However, when $\bar{P}_{h}$ departs from unity,  there would be
a sudden drop of the probability function if
the mass hierarchy is inverted:
$\bar{P}_{inv} \simeq \sin^{2} \phi_{13} \ll 1$, as shown in Fig. 3(b).
The sudden drop of this probability function 
is similar to that of $P_{nor}$ for the neutrino sector, and
can be characterized by the 
same condition for that of the neutrino, Eq.~(\ref{eq:g}),
with a slight change of the numerical values.
The properties of all the probability functions for $n > -4$ and
$n < -6$ are summarized in Table 1. 
 

\section{Power-law density profile, $\phi_{13}$, and mass hierarchies}


We shall now investigate whether and how the information about 
the mass hierarchy and $\phi_{13}$ could be extracted
from the observation of supernova neutrinos.  
Although the uncertainty in the density profile is unavoidable
due to incomplete knowledge of the supernova mechanism,
the results in Fig. 2 and Fig. 3 may still provide a useful guideline.  
We summarize the hints as follows.
 
\begin{enumerate} 
\item As suggested by Figs. 2(a) and 2(b), 
the mass hierarchy may be identified as normal if 
the survival probability of $\nu_{e}$ 
is observed to be extremely small, $i.e.$, if $P \ll 0.3$.
On the other hand, an extremely small survival probability
for $\bar{\nu}_{e}$, $\bar{P} \ll 0.7$, 
would signal an inverted hierarchy, as suggested by Figs. 3(a) and 3(b).
Although this feature provides no information about the mixing angle $\phi_{13}$,
it predicts the mass hierarchy without knowing the details of the density shape,
{\em i.e.}, the exact value of $n$ is irrelevant.
Note that the numerical values would vary slightly with the input parameters.  
\item If $P \simeq \bar{P} \approx 0.6$ is observed,
the power $n$ of the density profile is limited to $n < -6$ and a 
significant deviation from $n=-3$ is implied.  
The information about mass hierarchy or $\phi_{13}$ is
unavailable in this case. 
\item If $\bar{P} < P$ can be deduced from experiments,
then $n > -4$, $G_{h}(n,\phi_{13}) >1$, 
and the inverted mass hierarchy are implied.
These conditions lead to $\sin^{2}\phi_{13} > 4 \times 10^{-4}$.
On the other hand, if $\bar{P} > P$ is observed, it implies that
(i) the mass hierarchy is normal and $\phi_{13}$ is undetermined
(Figs. 2(a) and 3(a)), 
or (ii) the mass hierarchy is inverted and $G_{h}(n,\phi_{13}) <1$
(Figs. 2(b) and 3(b)).
A further observation of the Earth matter effect may be 
useful in selecting the correct scenario.
\item From Figs. 2(a) and 3(b), it is seen that $\phi_{13}$ can be sensitive to  
the direct observation of $P_{nor}$ or $\bar{P}_{inv}$ in only part of the parameter space. 
As discussed earlier, the prediction of $\phi_{13}$ depends crucially on the 
exact value of $n$ in this part of parameter space.
Thus, it is very difficult to establish a satisfactory constraint on $\phi_{13}$
from a direct observation of $P$ or $\bar{P}$ alone. 
For example,
Eq.~(\ref{eq:g}) suggests that a deviation of $\pm 0.5$ from $n=-3$ could result in an 
uncertainty in the prediction of $\phi_{13}$ by up to two orders of magnitude. 
A better knowledge of $n$ or 
an extended analysis that includes the Earth matter effect would help set
the constraint of this tiny mixing angle.  
\end{enumerate}
 
 
\section{Observation of Earth matter effects}

The regeneration effect of the supernova neutrinos when crossing the Earth has 
been widely discussed~\cite{dighe:00,fogli:02,luna:03,bandy:03,Ioan:05,ak:02}.
Since the Earth matter density, $\rho_{E} \sim$ a few $g/cm^{3}$,
is roughly the same order of magnitude as the density for the lower
level crossing in a supernova, the neutrino
fluxes may receive sizable modification due to the oscillation effects
in Earth.
In general, the Earth matter effect is signaled by 
the flux difference observed at two terrestrial 
detectors $D^{(1)}$ and $D^{(2)}$~\cite{dighe:00}:
\begin{equation}\label{eq:enor}
f_{e}^{(1)}-f_{e}^{(2)} \simeq P_{h} (1-2P_{l})
(P_{2e}^{(1)}-P_{2e}^{(2)})(f_{e}^{0}-f_{x}^{0}),
\end{equation}
\begin{equation}\label{eq:aenor}
f_{\bar{e}}^{(1)}-f_{\bar{e}}^{(2)} \simeq (1-2\bar{P}_{l})
(P_{1e}^{(1)}-P_{1e}^{(2)})(f_{\bar{e}}^{0}-f_{\bar{x}}^{0}),
\end{equation}
for the normal mass hierarchy, and
\begin{equation}\label{eq:einv}
f_{e}^{(1)}-f_{e}^{(2)} \simeq (1-2P_{l})
(P_{2e}^{(1)}-P_{2e}^{(2)})(f_{e}^{0}-f_{x}^{0}),
\end{equation}
\begin{equation}\label{eq:aeinv}
f_{\bar{e}}^{(1)}-f_{\bar{e}}^{(2)} \simeq P_{h} (1-2\bar{P}_{l})
(P_{1e}^{(1)}-P_{1e}^{(2)})(f_{\bar{e}}^{0}-f_{\bar{x}}^{0}),
\end{equation}
for the inverted mass hierarchy.  In the above expressions, 
$f_{e}^{(i)}$ ($f_{\bar{e}}^{(i)}$) is the observed 
$\nu_{e}$ ($\bar{\nu}_{e}$) flux at the detector $D^{(i)}$, 
$P_{je}^{(i)}$ ($\bar{P}_{je}^{(i)}$) is the probability that a 
$\nu_{j}$ ($\bar{\nu}_{j}$) arriving at the Earth surface is detected
as a $\nu_{e}$ ($\bar{\nu}_{e}$) at the detector, 
and $f^{0}_{e,x}$ ($f^{0}_{\bar{e},\bar{x}}$) is the initial flux
for the specific neutrino (antineutrino), with $x=\mu, \tau$.

The Earth matter effects could affect 
(i) $\nu_{e}$ flux only; (ii) $\bar{\nu}_{e}$ flux only;
(iii) both $\nu_{e}$ and $\bar{\nu}_{e}$ fluxes.  
We assume that the suppression of matter effect, if any,
is due solely to the smallness of $P_{h}$. 
Possible consequences of the above three scenarios are summarized in Table II
and discussed below: 

\begin{enumerate}
\item If the Earth effect is observed only in the $\nu_{e}$ flux
($f_{e}^{(1)}-f_{e}^{(2)} \neq 0$ and 
$f_{\bar{e}}^{(1)}-f_{\bar{e}}^{(2)} = 0$),
it requires an extremely small $P_{h}$ under 
the inverted hierarchy, as can be seen from Eqs.~(\ref{eq:einv})
and ~(\ref{eq:aeinv}).
Results in Table I show that
$P_{h} \rightarrow 0$ is possible when $n > -4$ and $G_{h}(n,\phi_{13})>1$.
These conditions lead to a lower bound on $\phi_{13}$:
$\sin^{2} \phi_{13} > 4 \times 10^{-4}$. 
\item If the Earth effect is observed only in the $\bar{\nu}_{e}$ flux
($f_{e}^{(1)}-f_{e}^{(2)} = 0$ and 
$f_{\bar{e}}^{(1)}-f_{\bar{e}}^{(2)} \neq 0$), then
Eqs.~(\ref{eq:enor}) and ~(\ref{eq:aenor})
suggest that $P_{h}$ is extremely small and the mass hierarchy is normal.
This leads to the same constraint:
$\sin^{2} \phi_{13} > 4 \times 10^{-4}$. 
\item If the Earth matter effect is observed in both the $\nu_{e}$ and 
the $\bar{\nu}_{e}$ fluxes, then $P_{h} \neq 0$ is required.  
It can be seen from Table I that this condition can be satisfied if
(i) $n > -4$ and $G_{h}(n,\phi_{13}) <1$, or (ii) $n < -6$.
Note that, as discussed earlier, the observation of supernova neutrino
loses its predictive power if $n < -6$.
Even from the first condition: $n > -4$ and $G_{h}(n,\phi_{13}) <1$,
an evident constraint on $\phi_{13}$ is still not available  unless the 
uncertainty of $n$ can be reduced significantly.
Furthermore, this scenario provides no information about the mass hierarchy.   
\end{enumerate}

There is a possibility that the constraint on $\phi_{13}$
might be available from checking the signs of $f_{e}^{(1)}-f_{e}^{(2)}$ and 
$f_{\bar{e}}^{(1)}-f_{\bar{e}}^{(2)}$, if the Earth matter effect is
observed in both the $\nu_{e}$ and the $\bar{\nu}_{e}$ fluxes. 
Suppose that 
one of the two detectors, $D^{(2)}$, is not shielded by the Earth matter.
It then leads to the replacements: $P_{2e}^{(2)} \rightarrow |U_{e2}|^{2}$ and
$\bar{P}_{1e}^{(2)} \rightarrow |U_{e1}|^{2}$, which satisfy
$P_{2e}^{(1)} \geq  |U_{e2}|^{2}$ and 
$\bar{P}_{1e}^{(1)} \geq |U_{e1}|^{2}$.
Since the average energies obey the
hierarchy: $\langle E(\nu_{e})\rangle < \langle E(\nu_{x}) \rangle$
and $\langle E(\bar{\nu}_{e})\rangle < \langle E(\bar{\nu}_{x}) \rangle$, 
there exists an energy $E_{c}$ ($\bar{E}_{c}$)
at which $f_{e}^{0}-f_{x}^{0}$ ($f_{\bar{e}}^{0}-f_{\bar{x}}^{0}$)
changes sign~\cite{dighe:00}. In general, $f_{e}^{0}-f_{x}^{0}>0$ 
($f_{\bar{e}}^{0}-f_{\bar{x}}^{0}>0$) in $E<E_{c}$ ($E<\bar{E}_{c}$),
and $f_{e}^{0}-f_{x}^{0}<0$ 
($f_{\bar{e}}^{0}-f_{\bar{x}}^{0}<0$) in $E>E_{c}$ ($E>\bar{E}_{c}$).
Furthermore,
the magnitude of $P_{l}$ varies from 0 (for $n > -4$) to 
$\cos^2 \theta_{12} \approx 0.7$ (for $n < -6$), and
$\bar{P}_{l}$ varies from 0 to $\sin^2 \theta_{12} \approx 0.3$.
Thus, $1-2\bar{P}_{l}$ is always positive, while $1-2P_{l}$ 
flips sign over the transition region $-6<n<-4$.  
The above arguments suggest the following: 
\begin{enumerate}
\item If $f_{e}^{(1)}-f_{e}^{(2)}$ and 
$f_{\bar{e}}^{(1)}-f_{\bar{e}}^{(2)}$ are both observed to be negative
at the high energy end of the spectrum, or both positive at the low energy end, 
then it implies $1-2P_{l}>0$ (from Table I),
$n > -4$, and $G_{h}(n,\phi_{13})>1$. 
This leads to the constraint: $\sin^{2} \phi_{13} > 4 \times 10^{-4}$. 
However, the mass hierarchy is undetermined from this result.
\item If $f_{e}^{(1)}-f_{e}^{(2)}$ and 
$f_{\bar{e}}^{(1)}-f_{\bar{e}}^{(2)}$ are of opposite signs,
then $1-2P_{l}<0$, and $n < -6$.  No further information about
$\phi_{13}$ or the mass hierarchy is available.
\end{enumerate}
We summarize the above results in Table III.

\section{Summary and conclusion}

The supernova neutrinos may
provide a promising future for the study of unknown neutrino properties. 
However, the detailed knowledge of the core-collapse supernova event
is still far from complete.  In addition to the 
uncertainties in the original neutrino fluxes and 
in the effects due to the shock propagation,
the original neutrino spectra can be further deformed by the
flavor conversion when the neutrinos propagate through
matter of uncertain density profiles.

In this work, parameters obtained from recent experiments are taken as
the input for the purpose of analyzing the survival probabilities of $\nu_{e}$
and $\bar{\nu}_{e}$. It is suggested that 
the influence coming from the energy variation can be excluded.
The effort is then focused on investigating how the unknown mass hierarchy, 
the mixing angle $\phi_{13}$, and the uncertainty in $n$ would affect the
probability functions.

It is shown that the non-trivial behavior of the probability functions can be well 
illustrated by the 3D plots in the $n-\sin^{2}\phi_{13}$ parameter space, and
that the uncertainty of $n$ could lead to
ambiguity in the interpretation of $\phi_{13}$ and the mass hierarchy. 
Roughly speaking, the probability functions behave
differently in three regions of the parameter space: 
$n < -6$, $-6 < n < -4$, and
$n > -4$.  As far as the mass hierarchy and the mixing angle
$\phi_{13}$ are concerned, the information is lost if the supernova neutrinos encounter 
a relatively steep density profile ($n < -6$) near the location
of flavor conversion.  
For a not as steep density profile ($-6 < n < -4$),
all the probability functions 
go through a transition that is governed by the variation of 
$n$.  This transition depends only very weakly on the mass hierarchy and $\phi_{13}$.
For $n > -4$, the probability functions vary with
the mass hierarchy, the value of $n$, and $\phi_{13}$ in a non-trivial fashion, 
as depicted clearly by Figs. 2 and 3.  Furthermore, this non-trivial structure
is found to be divided by a function of $n$ and $\phi_{13}$ through
the condition Eq.~(\ref{eq:g}).

For the qualitative observation of the
Earth matter effect, it can be shown that the constraint on
$\phi_{13}$ would be available only if $n > -4$.
However, the exact value of $n$ is irrelevant to the constraint 
as long as $n$ is greater than $-4$.

It is hoped that Eq.~(\ref{eq:g}) and
the 3D plots of the probability functions could 
provide a guideline to finding useful observables
from the future supernova neutrino experiments,
and to better help shed light on the desired  
understanding of the neutrino properties.
We shall return to this topic in the near future.


\acknowledgments S. -H. C. is supported by the National 
Science Council of Taiwan under the grant no. 94-2112-M-182-004.  
T. K. K. is supported in part by the DOE, 
grant no. DE-FG02-91ER40681.



 \begin{table}
 \begin{center}
 \begin{tabular}{lll}  
     & $n > -4$ & $n < -6$  \\ \hline
$P_{h} (\bar{P}_{h})$ & $  0$, if $G_{h}(n,\phi_{13}) >1$    &  $\cos^{2} \phi_{13}  \approx 1$  \\ 
                      & $ 1$, if $G_{h}(n,\phi_{13})<1$ &  \\ 
$P_{l}$ & 0   &  $\cos^{2} \theta_{12} \approx 0.7$ \\ 
 $\bar{P}_{l}$ & 0   &  $\sin^{2} \theta_{12} \approx 0.3$  \\   \hline
 $P$  & normal: $\sin^{2} \theta_{12} P_{h}+\sin^{2}\phi_{13} (1-P_{h})$  &      
  $\sin^{2} \theta_{12} +(\cos^{2}\theta_{12}-\sin^{2}\theta_{12})\cos^{2}\theta_{12} \approx 0.6$  \\ 
         & $\, \, \, \, \, \simeq \sin^{2}\phi_{13} \approx 0$, if $G_{h}(n,\phi_{13}) >1$ &   \\ 
         &  $\, \, \, \, \, \simeq \sin^{2}\theta_{12} \approx 0.3$, if $G_{h}(n,\phi_{13})<1$   & \\
         & inverted: $\sin^{2} \theta_{12} \approx 0.3$ &    \\  \hline
 $\bar{P}$ & normal: $\cos^{2}\phi_{13}\cos^{2}\theta_{12} \approx 0.7 $ & 
 $\cos^{2}\theta_{12}+(\sin^{2}\theta_{12} -\cos^{2} \theta_{12}) \sin^{2}\theta_{12} \approx 0.6$ \\
        & inverted: $\cos^{2}\theta_{12} P_{h}+\sin^{2}\phi_{13} (1-P_{h})$    & \\
        &  $\, \, \, \, \, \simeq \sin^{2}\phi_{13} \approx 0$, if $G_{h}(n,\phi_{13})>1$   &  \\
        & $\, \, \, \, \, \simeq \cos^{2}\theta_{12} \approx 0.7$, if $G_{h}(n,\phi_{13})<1$   & \\
     \end{tabular}
    \caption{Properties of the probability functions for $n > -4$
    and $n < -6$.}
  \end{center}
 \end{table}
 
\begin{table}
 \begin{center}
 \begin{tabular}{ccc}  
     matter effect & requirement &
       prediction   \\ \hline
      $\nu_{e}$ only   & $P_{h} \rightarrow 0$, $n > -4$, $G_{h}(n,\phi_{13}) >1$
        &  inverted, $\sin^{2} \phi_{13} > 4 \times 10^{-4}$   \\ 
      $\bar{\nu}_{e}$ only   & $P_{h} \rightarrow 0$, $n > -4$, $G_{h}(n,\phi_{13}) >1$  
      &  normal, $\sin^{2} \phi_{13} > 4 \times 10^{-4}$   \\ 
      both $\nu_{e}$ and $\bar{\nu}_{e}$   & $P_{h} \neq 0$  &  *   \\    
     \end{tabular}
    \caption{Predicting the mass hierarchy and $\phi_{13}$ from possible
    scenarios of the Earth matter effects. 
    (* See Table III for further predictions.)  }
  \end{center}
 \end{table}

\begin{table}
 \begin{center}
 \begin{tabular}{ccccc} 
     & $\bar{f}$ &
       $f$ &
       requirement
       & prediction  \\ \hline
      $E>E_{c}, (E>\bar{E}_{c})$   & -  &  +  &  $f_{e}^{0}-f_{x}^{0} <0$, $1-2P_{l} <0$
      & $n < -6$  \\ 
      (high energy end) &   &  -  &  $f_{e}^{0}-f_{x}^{0} <0$, $1-2P_{l} > 0$ &
       $n > -4$, $\sin^{2}\phi_{13} > 4 \times 10^{-4} $ \\  \hline
      $E<E_{c}, (E<\bar{E}_{c})$   & +  &  +  &  $f_{e}^{0}-f_{x}^{0} >0$, $1-2P_{l} > 0$ &
       $n > -4$, $\sin^{2}\phi_{13} > 4 \times 10^{-4}$  \\  
       (low energy end) &   &  - & $f_{e}^{0}-f_{x}^{0} >0$, $1-2P_{l} < 0$ 
        &$n < -6$  \\ 
     \end{tabular}
    \caption{Predicting $\phi_{13}$ and $n$ from the signs of 
    $\bar{f} \equiv f^{(1)}_{\bar{e}}-f^{(2)}_{\bar{e}}$
    and $f \equiv f^{(1)}_{e}-f^{(2)}_{e}$ if
    the Earth matter effect is observed in both $\nu_{e}$ and $\bar{\nu}_{e}$
    fluxes.  Note that the information about the mass hierarchy is unavailable. }
  \end{center}
 \end{table} 


\begin{figure}
\centerline{\epsfig{figure=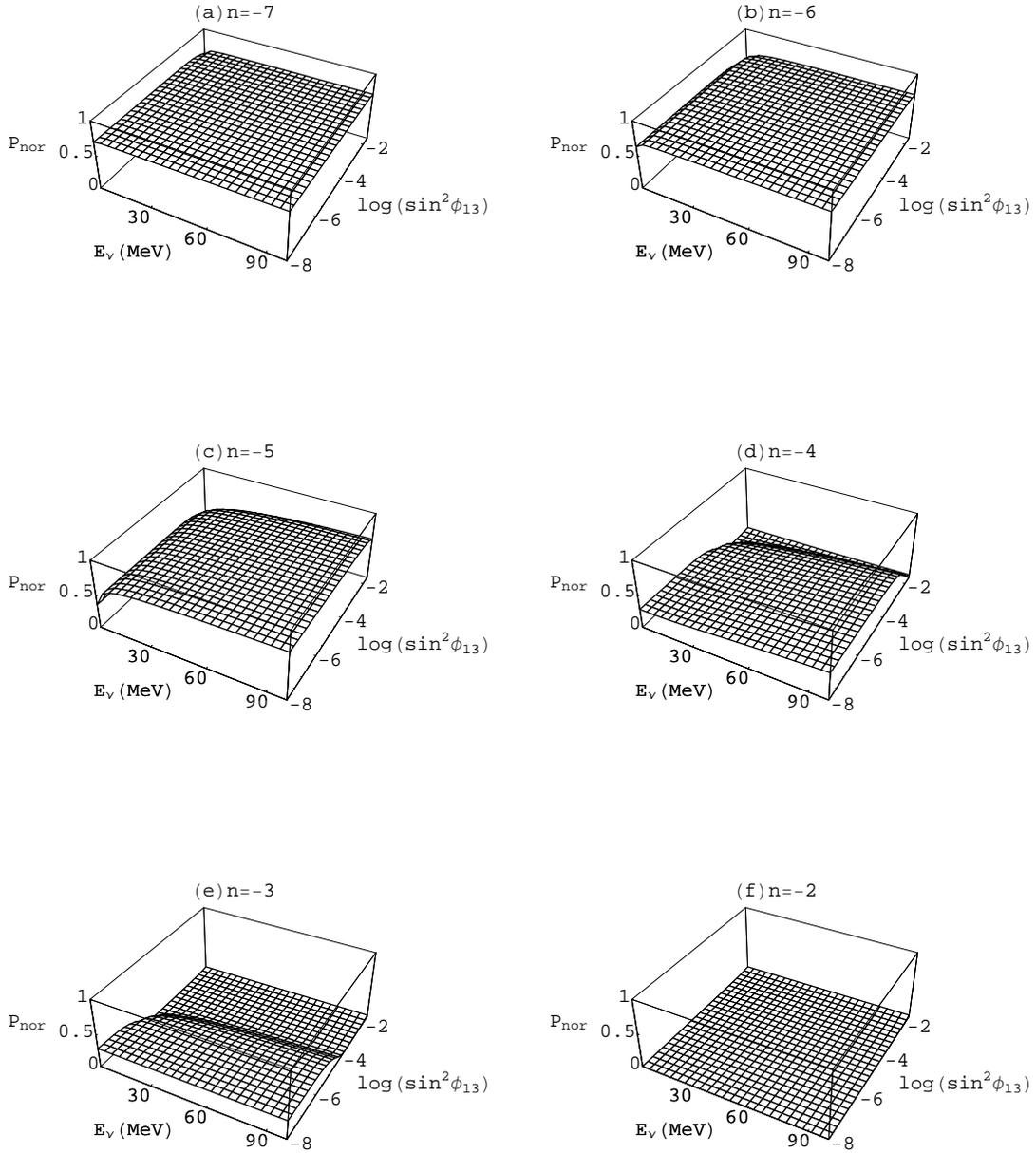,width=6in}}  
\caption{The 3D plots of $P_{nor}=P_{nor}(E_{\nu},\phi_{13})$
for different values of $n$. 
The following values are adopted: $\Delta m^{2}_{32}=3.0 \times 10^{-3}$eV$^{2}$,
$\Delta m^{2}_{21}=7.0 \times 10^{-5}$eV$^{2}$, $\sin^{2}\theta_{12} =0.8$, 
and $c=7.0 \times 10^{31}$ g $\cdot$ cm$^{n-3}$.} 
  \label{Figure 1}
    \end{figure} 
    

\begin{figure}
\centerline{\epsfig{figure=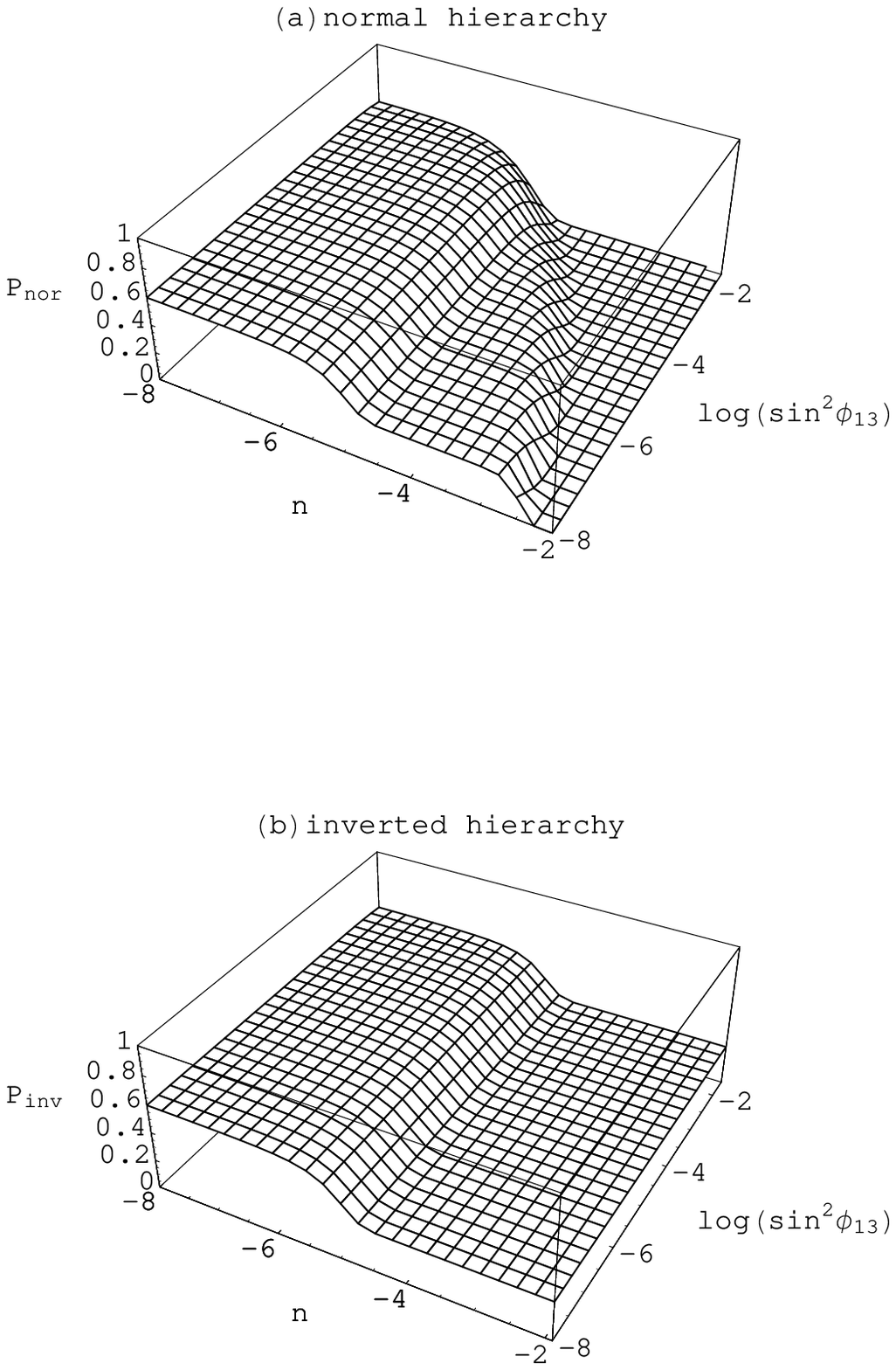,width=5in}}  
\caption{The 3D plots of $P=P(n,\phi_{13})$ 
under both (a) the normal, and (b) the inverted mass hierarchies.  The average energy 
$\langle E_{\nu} \rangle =12$ MeV is adopted.}
  \label{Figure 2}
    \end{figure} 


\begin{figure}
\centerline{\epsfig{figure=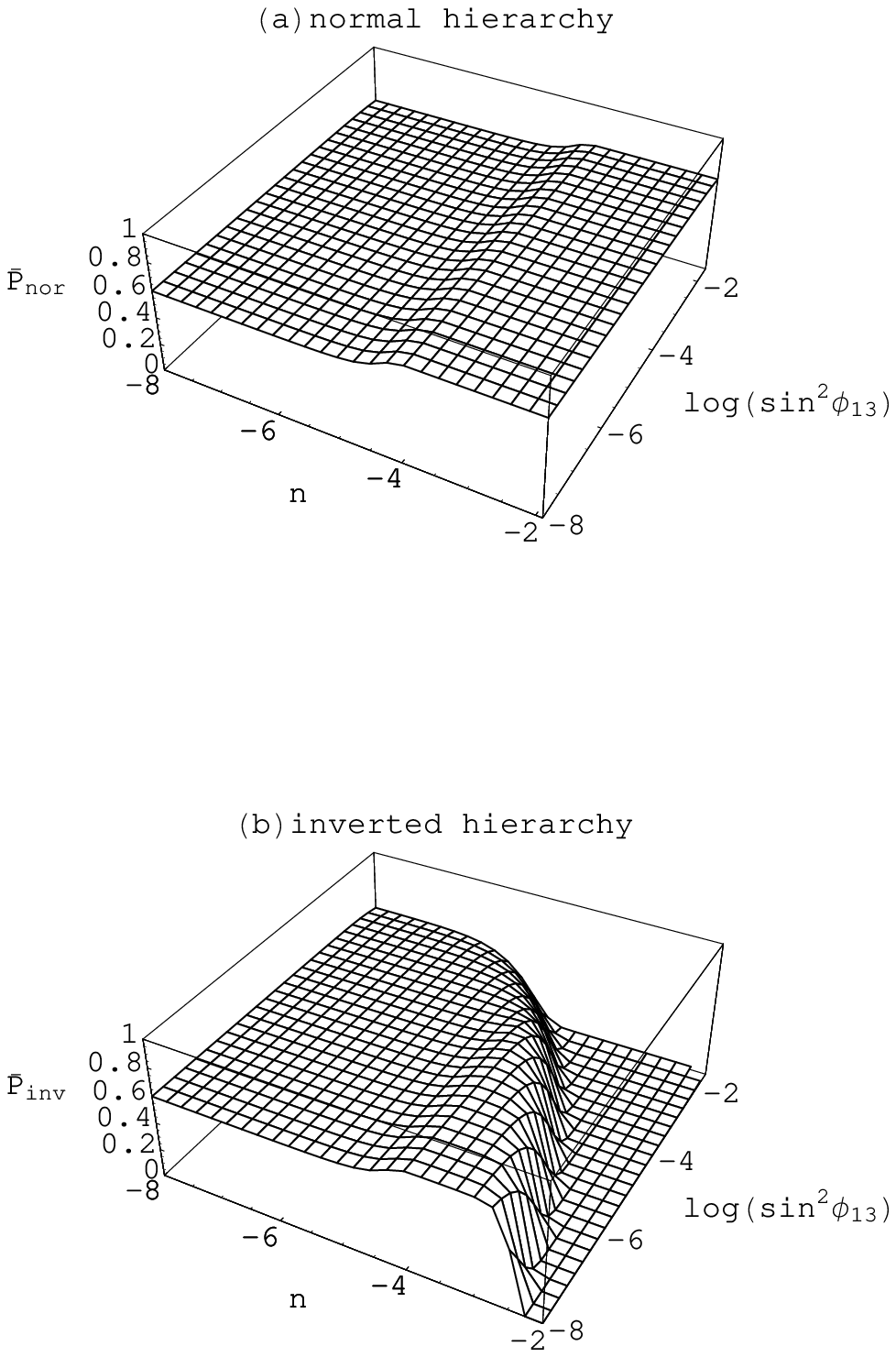,width=5in}}  
\caption{The 3D plots of $\bar{P}=\bar{P}(n,\phi_{13})$ 
under both (a) the normal and (b) the inverted mass hierarchies. The average energy 
$\langle E_{\bar{\nu}} \rangle =15$ MeV is adopted.}
  \label{Figure 3}
    \end{figure} 

\end{document}